\documentclass[aps,pra,twocolumn,showpacs,letterpaper,superscriptaddress]{revtex4-1}

\usepackage[colorlinks=true, citecolor=blue, linkcolor=blue, urlcolor=blue]{hyperref}

\usepackage{graphicx,dcolumn,longtable,epsfig}
\usepackage[usenames]{color}
\usepackage{amssymb}
\usepackage{amsmath}
\usepackage{epstopdf}
\usepackage{bm}
\usepackage{footnote}
\usepackage{float}
\usepackage{subfigure}
\usepackage{color}
\usepackage{ulem}
\usepackage[version=4]{mhchem}
\usepackage[T1]{fontenc}

\input epsf.tex

\def\bea{\begin{eqnarray}}
\def\eea{\end{eqnarray}}
\def\be{\begin{equation}}
\def\ee{\end{equation}}

\begin{document}
\title{
Ground states of 2D tilted dipolar bosons with density-induced hopping}
\author{Chao Zhang}
\email{chaozhang@umass.edu}
\affiliation{State Key Laboratory of Precision Spectroscopy, East China Normal University, Shanghai 200062, China}
\affiliation{Department of Physics, University of Massachusetts, Amherst, Massachusetts 01003, USA}
\author{Jin Zhang}
\email{jin-zhang@uiowa.edu}
\affiliation{Department of Physics and Astronomy, University of Iowa, Iowa City, IA 52242, USA}
\author{Jin Yang}
\affiliation{Department of Physics, University of Virginia, Charlottesville, Virginia 22904, USA}
\author{Barbara Capogrosso-Sansone}
\affiliation{Department of Physics, Clark University, Worcester, Massachusetts 01610, USA}

\begin{abstract}
Motivated by recent experiments with ultracold magnetic atoms trapped in optical lattices where the orientation of atomic dipoles can be fully controlled by external fields, we study, by means of quantum Monte Carlo, the ground state properties of dipolar bosons trapped in a two-dimensional lattice with density-induced hopping and where the dipoles are tilted along the $xz$ plane. We present ground state phase diagrams of the above system at different tilt angles.
We find that, as the dipolar interaction increases, the superfluid phase at half filling factor is destroyed in favor of either a checkerboard or stripe solid phase for tilt angle $\theta \lesssim 30^{\circ}$ or $\theta \gtrsim 30^{\circ}$ respectively. More interesting physics happens at tilt angles $\theta \gtrsim 58^{\circ}$ 
where we find that, as the dipolar interaction strength increases,  solid phases   first appear at filling factor lower than $0.5$. Moreover, unlike what observed at lower tilt angles, we find that, at half filling, a stripe supersolid intervenes between the superfluid and stripe solid phase.
\end{abstract}

\maketitle

\section{Introduction}
\label{sec:sec1}

Long-range interactions have attracted a great deal of attention in the cold atom community as  they have been theoretically shown to stabilize a plethora of exotic quantum phases \cite{Goral:2002hu}.
State-of-the-art ultracold experiments have paved the way to experimentally explore these exotic phases. 
Long-range interactions can be realized with ultracold atoms with large magnetic moments~\cite{Baier:2016ga,Griesmaier:2005fd, DePaz:2013ff}, by exciting atoms into Rydberg states~\cite{Schauss:2015ch,Bernien2017}, or with ultracold dipolar molecules~\cite{Yan:2013fn, Hazzard:2014bx,Frisch:2015gm,Seesselberg:2018ff,Lu:2012bd,Lu:2011hl}. The latter can realize strong dipolar long-range interactions which may demand an extension or a revision of the standard Bose-Hubbard model~\cite{CapogrossoSansone:2010em,Sowinski:2012kl, Maik:2013ji,Biedron:2018iz}. Another way to realize long-range interactions is implementing an optical lattice in an optical cavity, where the long-range interaction is mediated by strong matter-light interaction inside the cavity~\cite{Baumann:2010js, Landig:2016il}. This possibility has triggered extensive theoretical endeavors into long-range interaction mediated by optical cavities~\cite{Sundar:2016ie, Habibian:2013eh, Dogra:2016hy, Flottat:2017gn, Zhang:2020ko, Habibian:2013kw, Zhang:2020eo}. Other theoretical notable works focused on long-range interactions in the presence of disorder or doping~\cite{Zhang:2017ei, Grimmer:2014kx}, which can be easily implemented in ultracold experiments. Since optical lattices are remarkablely versatile compared to solid state experiments, there have been several pioneering theoretical works considering long-range interactions in optical lattices with anisotropic tunneling rates~\cite{Lingua:2018dn,SafaviNaini:2014kc}. These works so far are limited to multiple layers of one dimensional optical lattices due to large computational resources needed with the increase of the number of nearest neighbors in three-dimensions. In ultracold experiments, both strengths and directions of magnetic or electric field can be freely tuned. As a result, the direction of dipoles can be adjusted freely. In recent years, many efforts were put into exploring physics of long-range dipolar interactions with different tilt angles~\cite{Danshita:2009cp,Wu:2020kf,Bandyopadhyay:2019ew,Zhang:ws,Bhongale:2012fb,Parish:2012gq,PhysRevA.82.013643,PhysRevLett.109.235307,PhysRevA.84.063633}.

Dipolar interactions are anisotropic. When two dipoles are placed side by side, they repel each other; when they are placed head to tail, they attract each other. Most of the early studies on dipolar interactions focused on systems with dipoles aligned perpendicular or parallel to the lattice plane. Tilted dipolar interactions with arbitrary angles have been theoretically studied in ultracold gases systems without lattice potentials (thus not based on Hubbard model)~\cite{PhysRevA.82.013643, Sun:2010go, Block:2012hr, Parish:2012gq,PhysRevLett.109.235307,PhysRevA.84.063633}, or in lattices using renormalization group~\cite{Bhongale:2012fb}, mean field theory~\cite{Danshita:2009cp,Wu:2020kf}, and variational approaches~\cite{Goral:2002hu}.  In reference~\cite{Zhang:ws}, the authors used quantum Monte Carlo method and found the ground state phase diagram as a function of tilt angle $\theta$ at half filling and for hard-core bosons. For soft-core bosons, the ground state phase diagrams have been found for tilt angles in the range $0\le \theta\le45^{\circ}$~\cite{Bandyopadhyay:2019ew}.
The above mentioned studies do not consider density-induced hopping, 
and, more generally, the details of how the parameters tuned experimentally, i.e. the scattering length, the dipole moment, and the depth of the optical lattice potential affect the onsite interaction, the long-range interaction, and the hopping strength entering the effective model used to describe the system. A recent experiment~\cite{Baier:2016ga} has realized dipolar bosons in a three-dimensional lattice and considered how all the experimentally tunable parameters, such as scattering length and dipolar interaction strength, affect onsite and long-range interaction, and hopping. This experiment paves the way to investigate quantum phase transitions of tilted dipolar lattice bosons with density-induced hopping.

In this paper, we use path-integral Monte Carlo simulations based on the worm algorithm~\cite{Prokofev:1998gz} to study the ground state phase diagram of dipolar bosons in a two-dimensional lattice with density-induced hopping where the dipoles are tilted on the $xz$ plane. In the absence of sign-problem, path-integral Monte Carlo in continuous time is approximation-free and produces unbiased results, that is, errors are controllable and purely statistical. 
We calculate the parameters entering the effective model, i.e. the onsite interaction, long-range interaction strength, and density-induced hopping from the parameters that can be tuned experimentally, such as the scattering length, dipolar interaction strength, and optical lattice potential depth. The paper is organized as follows: in section~\ref{sec:sec2}, we introduce the Hamiltonian of the system and the parameters that can be controlled in experiments. In section~\ref{sec:sec3}, we discuss various phases and the corresponding order parameters. In section~\ref{sec:sec4}, we present the phase diagrams of the above system at four tilt angles and discuss the nature of the transitions. In section~\ref{sec:sec5}, we briefly discuss the experimental realization. We conclude the article in section~\ref{sec:sec6}. 

\section{Hamiltonian} 
\label{sec:sec2}

\begin{figure}[h]
\includegraphics[trim=0.5cm 6cm 1cm 3cm, clip=true, width=0.48\textwidth]{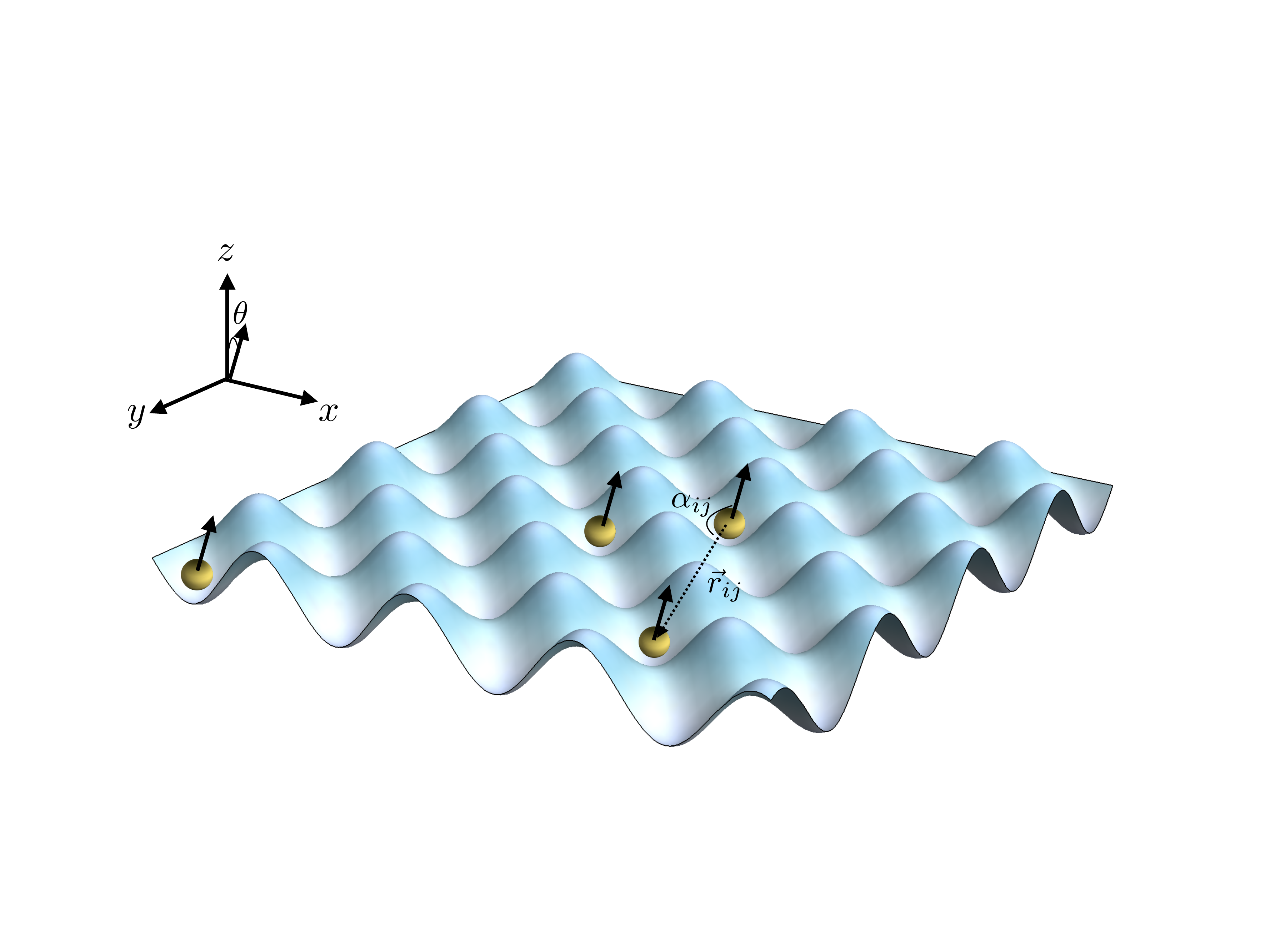}
\caption{Schematic representation of the system. Dipoles are trapped in a two-dimensional optical lattice and are aligned parallel to
each other along the direction of polarization, determined by an electric/magnetic field. $\theta$ is the angle between polarization and $z$ direction, $\vec{r}_{\mathbf{ij}}$ is the relative position between site $\mathbf{i}$ and $\mathbf{j}$. $\alpha_{\mathbf{ij}}$ is the angle between polarization and $\vec{r}_{\mathbf{ij}}$.}
\label{FIG1}
\end{figure}

\begin{figure}[h]
\includegraphics[width=0.48\textwidth]{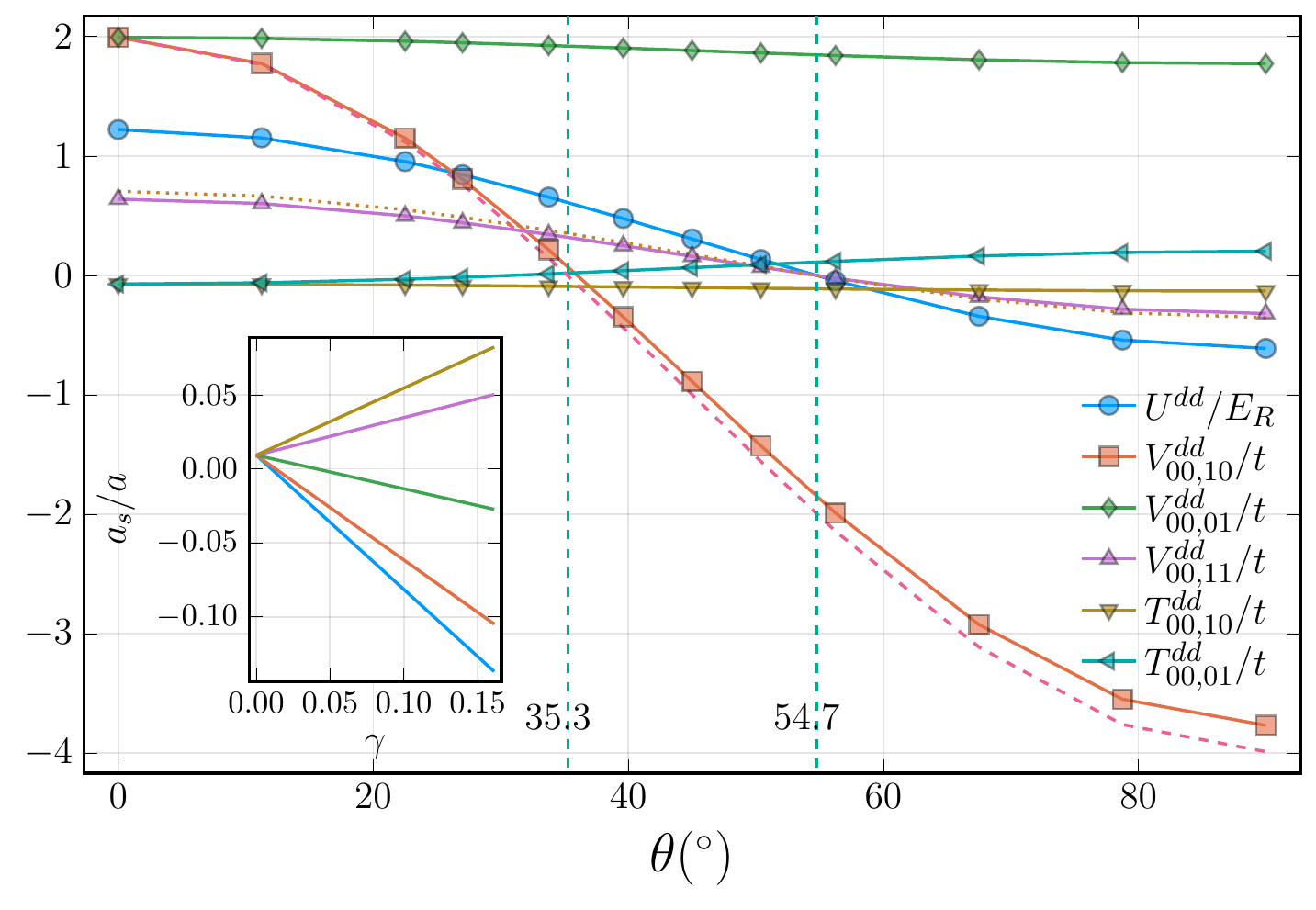}
\caption{Dipolar contribution to Hamiltonian parameters as a function of tilt angle $\theta$ at $\gamma = 1/\pi^3, s = \kappa = 10$. The two vertical dashed lines locate angles $35.3^{\circ}$ and $54.7^{\circ}$ respectively. The dashed line by $V^{dd}_{\mathbf{i}, \mathbf{i}+\hat{\mathbf{x}}}/t$ is $V(1-3sin^2(\theta))$. The dotted line by $V^{dd}_{\mathbf{i}, \mathbf{i}+\hat{\mathbf{x}}+\hat{\mathbf{y}}}/t$ is $V\left(1-3sin^2(\theta)cos^2(45^{\circ})\right) / (\sqrt{2})^3$. Lines in the inset show $a_s / a$ vs $\gamma$ at fixed onsite interaction $U/t = 20$ for $\theta = 0^{\circ}, 22.5^{\circ}, 45^{\circ}, 67.5^{\circ}, 90^{\circ}$ from bottom up, where $0 < \gamma < 0.16$ and $-0.137 < a_s / a < 0.082$. }
\label{fig:hamparamsvstheta}
\end{figure}

We study dipolar bosons with atomic mass $m$ in a square optical lattice created by a separable external potential $V_{\text{ext.}}(x,y,z) = V_0 \left[ cos^2(k_L x) + cos^2(k_L y) \right] + m \Omega^2_z z^2 / 2$. The laser beams with a wavelength $\lambda = 2a$ ($a$ the lattice spacing) in the $xy$ plane generate a two dimensional square lattice with $k_L = 2\pi/\lambda$  the lattice momentum. The lattice depth is written in units of recoil energy $V_0 = s E_R$. The recoil energy $E_R = \hbar^2 k_L^2 / 2m $ defines a natural energy scale of the system. $\Omega_z$ is the frequency of the harmonic trap in the $z$ direction, controlling the thickness of the two dimensional sheet, from which we define the lattice flattening constant $\kappa = \hbar \Omega_{z} / 2E_R$ \cite{Sowinski:2012kl}. As shown in Fig.~\ref{FIG1}, we assume all dipole moments to be in the same direction and to rotate in the $x z$ plane with tilt angle $\theta$ between the dipole moment and the $z$ axis. The many-body Hamiltonian describing this system in the second quantization language reads
\begin{eqnarray}
\label{eq:bosonham2ndquant}
\hat{H} &= \int d^3 \mathbf{r} \psi^{\dagger}(\mathbf{r})\left[-\frac{\hbar^{2} \nabla^{2}}{2 m}+V_{\text {ext. }}(\mathbf{r})\right] \psi(\mathbf{r}) \nonumber \\
&+\frac{1}{2} \int \psi^{\dagger}(\mathbf{r}) \psi^{\dagger}\left(\mathbf{r}^{\prime}\right) V\left(\mathbf{r}^{\prime}-\mathbf{r}\right) \psi\left(\mathbf{r}^{\prime}\right) \psi(\mathbf{r}) d^{3} \mathbf{r} d^{3} \mathbf{r}^{\prime},
\end{eqnarray}
where $\psi^\dagger$ ($\psi$) is the bosonic creation (annihilation) field operator. The interaction between dipolar bosons contains contact ($V_c$) and dipole-dipole ($V_{dd}$) interactions,
\begin{eqnarray}
\nonumber V(\mathbf{r}-\mathbf{r}^{\prime}) &=& V_c(\mathbf{r}-\mathbf{r}^{\prime}) + V_{dd}(\mathbf{r}-\mathbf{r}^{\prime}) \\ 
&=& \tilde g \delta(\mathbf{r}-\mathbf{r}^{\prime}) + \tilde\gamma \frac{1-3cos^2(\alpha)}{|\mathbf{r}-\mathbf{r}^{\prime}|^3}
\end{eqnarray}
with $\tilde g = 4\pi\hbar^2 a_s / m$, $a_s$ the $s$-wave scattering length, and $\tilde\gamma = \mu_e^2 / (4\pi \epsilon_0)$ or $\mu_0 \mu_m^2 / 4\pi$, $\mu_e$ ($\mu_m$) the electric (magnetic) dipole moment of bosons, $\epsilon_0$ ($\mu_0$) the vacuum permittivity (permeability). $\alpha$ is the angle between the direction of dipole moments and the relative position of two bosons. The bosonic field operator can be expanded with Wannier functions \cite{Kohn:1959bn} in the lowest Bloch band $\psi(\mathbf{r}) = \sum_{\mathbf{i}} W_\mathbf{i}(x,y,z) \hat{a}_{\mathbf{i}}$. One then arrives at the extended Bose-Hubbard (EBH) model
\begin{align}
\nonumber H& =-t\sum_{\langle \mathbf{i}, \mathbf{j}\rangle } a_\mathbf{i}^\dagger a_\mathbf{j} + \frac{U}{2}\sum_\mathbf{i} n_\mathbf{i}(n_\mathbf{i}-1) + \frac{1}{2} \sum_{ \mathbf{i},\mathbf{j} } V_{\mathbf{i},\mathbf{j}} n_\mathbf{i} n_\mathbf{j} \\
&-\sum_{\langle \mathbf{i}, \mathbf{j}\rangle}T_{\mathbf{i},\mathbf{j}}a_\mathbf{i}^{\dagger}(n_\mathbf{i}+n_\mathbf{j}) a_\mathbf{j}- \mu \sum_\mathbf{i} n_\mathbf{i} \;\; ,
\label{Eq1}
\end{align}
where the first term is the kinetic energy characterized by the hopping amplitude $t$. 
Here $\langle \cdots \rangle$ denotes nearest-neighboring sites and $a_\mathbf{i}^\dagger$ ($a_\mathbf{i}$) are bosonic creation (annihilation) operators satisfying the bosonic commutation relations $[a_\mathbf{i}, a_\mathbf{j}^\dagger] = \delta_{\mathbf{ij}}$. 
$U$ is the onsite repulsive interaction, and $n_\mathbf{i}=a_\mathbf{i}^{\dagger}a_\mathbf{i}$ is the particle number operator. 
$V_{\mathbf{ij}}$ is the off-site interaction between sites $\mathbf{i}$ and $\mathbf{j}$. If the lattice is deep enough, $V_{\mathbf{ij}} \approx V \left( 1 - 3cos^2(\alpha)\right) / |\mathbf{i}-\mathbf{j}|^3$ ($V$ is the nearest-neighbor interaction when dipole moments are along $z$ axis), which is widely used in EBH model to study phase transitions. For a perfect two dimensional system, $cos(\alpha) = sin(\theta)cos(\phi)$ with $\phi$ the polar angle of the relative position $\mathbf{i}-\mathbf{j}$ in the $xy$ plane, the off-site dipole-dipole interactions in $\hat{\mathbf{x}}$ and $\hat{\mathbf{x}}+\hat{\mathbf{y}}$ directions are negative for $\theta > sin^{-1}(1/\sqrt{3}) \approx 35.3^{\circ}$ and $\theta > sin^{-1}(\sqrt{2/3}) \approx 54.7^{\circ}$ respectively, while those in $y$ direction is positive and independent of $\theta$ because $\alpha = 90^{\circ}$. It can be shown that the contribution $U^{dd}$ from the dipolar interaction to the on-site strength of the EBH model  (see below) is zero at angle $54.7^{\circ}$ because of the rotational symmetry of Wannier functions.
$T$ is the density-induced tunneling. We also introduce the chemical potential in the last term to control the total number of bosons. We neglect the pair tunneling term because its strength is very small (see Appendix~\ref{apdx:calparams}).

All interaction parameters entering the Hamiltonian in Eq.~\ref{Eq1} can be found from Eq.~\ref{eq:bosonham2ndquant} by calculating integrals involving Wannier functions (see Appendix \ref{apdx:calparams}) in units of recoil energy and lattice coordinate $\mathbf{r} \rightarrow \mathbf{r} / a$, where $g = 8a_s/(\pi a), \gamma = m \mu_e^2 / (2\pi^3 \epsilon_0 \hbar^2 a)$ or $\mu_0 \mu_m^2 m / (2\pi^3\hbar^2 a)$. Contact interactions and dipolar interactions in Eq.~\ref{eq:bosonham2ndquant} give us two sets of parameters $U^c$, $V_{\mathbf{i}, \mathbf{j}}^c$, $T_{\mathbf{i}, \mathbf{j}}^c$ and $U^{dd}$, $V_{\mathbf{i}, \mathbf{j}}^{dd}$, $T_{\mathbf{i}, \mathbf{j}}^{dd}$ that determine the Hamiltonian parameters $U = U^c+U^{dd}, V = V^c+V^{dd}, T = T^c - T^{dd}$ \cite{Dutta:2015fe}. We consider  lattice depth $s = 10$ so that the tight binding approximation for contact interactions holds. To obtain a valid approximation of a two dimensional system, we need $\kappa \gg \sqrt{s}$ so that the energy gap in the $z$ direction is much larger than the one in the $xy$ plane. We use lattice flattening $\kappa = 10$. Then the tunneling $t = 0.0192E_{R}$ is fixed, where $E_{R}$ is the recoil energy, while the parameters obtained from contact interactions ($U^c$, $V_{\mathbf{i}, \mathbf{j}}^c$, $T_{\mathbf{i}, \mathbf{j}}^c$ ) are proportional to $g$ and those obtained from dipolar interactions ( $U^{dd}$, $V_{\mathbf{i}, \mathbf{j}}^{dd}$, $T_{\mathbf{i}, \mathbf{j}}^{dd}$) are proportional to $\gamma$. We calculate off-site interactions $V^{dd}_{\mathbf{i}, \mathbf{j}}$ for relative lattice positions $| \mathbf{i}- \mathbf{j} | \le 5$ to keep track of long-range effects of dipolar interactions, while we only keep nearest neighbor terms for hopping $t$ and density-induced hopping $T$ since the decay of hopping and density-induced hopping is exponentially fast and the long-range part can be neglected.

In Fig.~\ref{fig:hamparamsvstheta}, we show how the parameters obtained from the dipolar interaction depend on the tilt angle $\theta$ for $\gamma = 1/\pi^3$. Notice that the dipolar part of on-site interaction $U^{dd}$ is in units of recoil energy. As we increase the tilt angle, $U^{dd}$, the nearest-neighbor interaction in $\hat{\mathbf{x}}$ direction $V^{dd}_{\mathbf{i}, \mathbf{i}+\hat{\mathbf{x}}}$, and the next-nearest-neighbor interaction in $\hat{\mathbf{x}}+\hat{\mathbf{y}}$ direction $V^{dd}_{\mathbf{i}, \mathbf{i}+\hat{\mathbf{x}}+\hat{\mathbf{y}}}$ go from positive to negative; $T^{dd}_{\mathbf{i}, \mathbf{i}+\hat{\mathbf{y}}}/t$ goes from $-0.073$ to $0.203$; $T^{dd}_{\mathbf{i}, \mathbf{i}+\hat{\mathbf{x}}}/t$ is always negative from $-0.073$ to $-0.131$; $V^{dd}_{\mathbf{i}, \mathbf{i}+\hat{\mathbf{y}}}/t$ does not change much as expected. $V^{dd}_{\mathbf{i}, \mathbf{i}+\hat{\mathbf{x}}} = 0$ at $\theta \gtrsim  35.3^{\circ}$, and $U^{dd} = V^{dd}_{\mathbf{i}, \mathbf{i}+\hat{\mathbf{x}}+\hat{\mathbf{y}}} = 0$ at angle $54.7^{\circ}$. The widely used approximation $V_{\mathbf{i}, \mathbf{j}} \approx V \left( 1 - 3cos^2(\alpha)\right) / |\mathbf{i-j }|^3$ is also plotted for $V^{dd}_{\mathbf{i}, \mathbf{i}+\hat{\mathbf{x}}}$ (dashed line) and $V^{dd}_{\mathbf{i}, \mathbf{i}+\hat{\mathbf{x}}+\hat{\mathbf{y}}}$ (dotted line) ($V^{dd}_{\mathbf{i}, \mathbf{i}+\hat{\mathbf{y}}}$ in the approximation is a constant). We see that for $V^{dd}_{\mathbf{i}, \mathbf{i}+\hat{\mathbf{x}}}$ and $V^{dd}_{\mathbf{i}, \mathbf{i}+\hat{\mathbf{y}}}$, the approximation slightly deviates from the numerical results only at large angles, while it is good at all angles for $V^{dd}_{\mathbf{i}, \mathbf{i}+\hat{\mathbf{x}}+\hat{\mathbf{y}}}$. 
The agreement between calculated parameters and the approximation indicates that the 2D approximation is valid.  
In the following, we fix the onsite interaction $U/t = (U^c+U^{dd})/t = 20$ and study phase diagrams in the $\gamma$-$n$ plane for different tilt angles $\theta$, where $n$ is the filling factor. The inset of Fig.~\ref{fig:hamparamsvstheta} shows the dependence of $a_s / a$ on $\gamma$ with fixed $U/t = 20$. The largest $\gamma$ we use to calculate the phase diagram is $0.16$, and the $s$-wave scattering length in units of lattice spacing goes from $-0.137$ to $0.082$ as we increase the tilt angle from $0^{\circ}$ to $90^{\circ}$.


\section{Quantum phases and order parameters}
\label{sec:sec3}

\begin{figure}[h]

\includegraphics[trim=3cm 9cm 1cm 3cm, clip=true, width=0.52\textwidth]{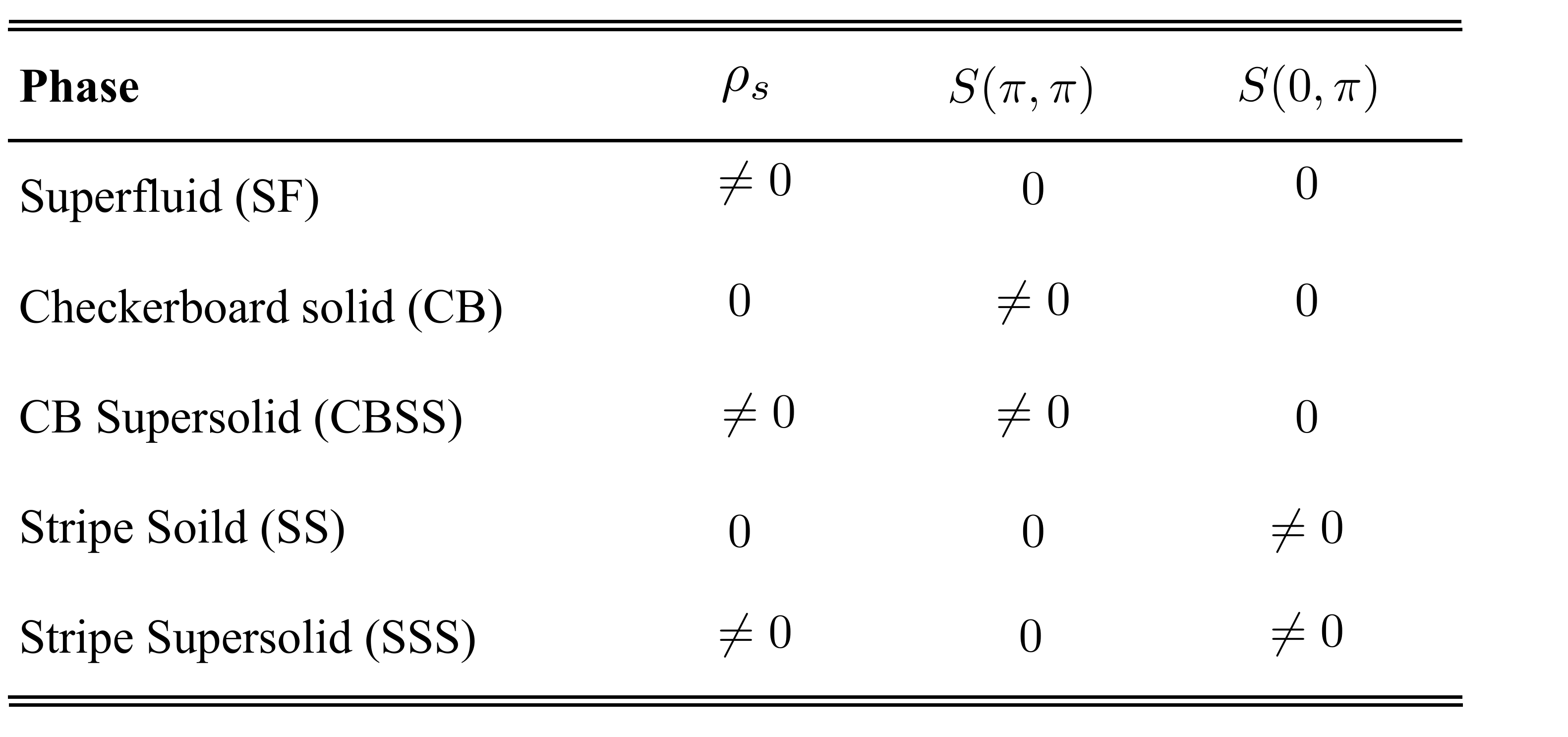}
\caption{Quantum phases and the corresponding order parameters: superfluid density $\rho_s$, structure factor $S(\pi, \pi)$, and $S(0,\pi)$.
}
\label{Table1}
\end{figure}

In this section, we list the phases stabilized by Eq.~\ref{Eq1} and the corresponding order parameters. Fig.~\ref{Table1} shows order parameters for superfluid (SF) phase, checkerboard solid (CB) phase, checkerboard supersolid (CBSS) phase, stripe solid (SS) phase, stripe supersolid (SSS) phase. 
Each phase corresponds to a unique combination of the order parameters. Here, three order parameters are needed in order to characterize the quantum phases: superfluid density $\rho_s$, structure factor $S(\pi, \pi)$, and $S(0, \pi)$.

The superfluid density is calculated in terms of the winding number~\cite{Winding}: $\rho_s=\langle \mathbf{W}^2 \rangle /dL^{D-2}\beta$, where $\mathbf{W}$ is the winding number and $\mathbf{W}^2=W_x^2+W_y^2$. 
 
The structure factor characterizes diagonal long-range order and is defined as: $S(\mathbf{k})=\sum_{\mathbf{r},\mathbf{r'}} \exp{[i \mathbf{k} \cdot (\mathbf{r}-\mathbf{r'})]\langle n_{\mathbf{r}}n_{\mathbf{r'}}\rangle}/N$, where N is the particle number. Here, $\mathbf{k}$ is the reciprocal lattice vector. We use $\mathbf{k}=(\pi, \pi)$ and $\mathbf{k}=(0,\pi)$ to identify the CB and SS phases respectively.

\section{Ground state phase diagrams}
\label{sec:sec4}

\begin{figure*}[th]
\centering
\includegraphics[trim=0.5cm 3cm 0cm 3cm, clip=true, width=1.0\textwidth]{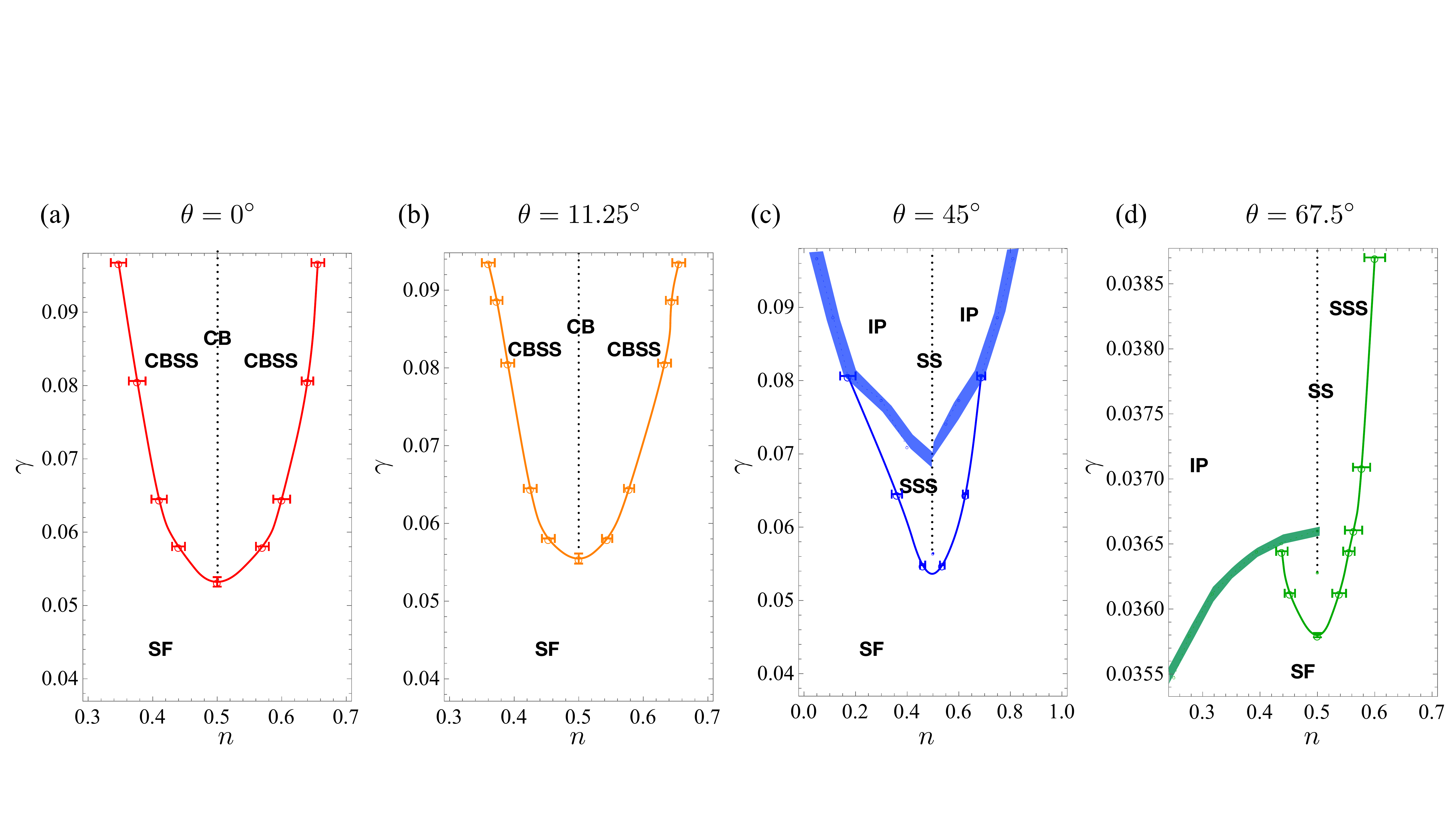}
\caption{Ground state phase diagram for $\theta=0^{\circ}$(a), $\theta=11.25^{\circ}$(b), $\theta=45^{\circ}$(c), and $\theta=67.5^{\circ}$(d). The x-axis is filling factor $n$ and the y-axis is the dipolar interaction strength $\gamma$. For tilt angles $\theta\lesssim 30^{\circ}$, the solid phase stabilized at half filling corresponds to a checkerboard solid (CB) and the supersolid phase is a CBSS; for $\theta\gtrsim30^{\circ}$, the solid phase corresponds to a stripe solid (SS) and the supersolid phase is a SSS. IP stands for the incompressible ground states stabilized at rational filling factors. Solid lines correspond to second-order transitions while solid shaded regions correspond to first order phase transitions. Dotted lines at filling factor $n=0.5$ represent solid phases CB or SS. When not visible, error bars are within symbol size.}
\label{FIG2}
\end{figure*}

In this section, we present ground state phase diagrams of dipolar bosons in a square lattice and with density-induced hopping. The dipoles are parallel to each other and are tilted in the $xz$ plane. Fig.~\ref{FIG2} shows the ground state phase diagram for four tilt angles $\theta=0^{\circ}$, $11.25^{\circ}$, $45^{\circ}$, and $67.5^{\circ}$ at fixed $U/t=20$. The $x$ axis is the filling factor $n=N/N_{site}$, where $N$ is the particle number and $N_{site}=L \times L$, with $L$ the system size. Here, we consider $n<1.0$. The $y$ axis is the dipolar interaction strength $\gamma$. The transition points on the phase diagrams are determined using system sizes $L=20$, 40, and 60 and inverse temperature $\beta=L$. This choice assures that temperature is low enough so that we are effectively at zero temperature and we are therefore probing ground-state properties. For second-order phase transitions we have performed standard finite-size scaling.

In Fig.~\ref{FIG2}(a) we plot the phase diagram for the system with all dipoles tilted perpendicularly to the $xy$ plane. If we compare with results in~\cite{Grimmer:2014kx}, where density-induced hopping is not taken into account, we see that superfluidity is slightly suppressed. This is due to negative density-induced hopping ($T_{x}\sim -0.17$ and $T_y\sim -0.17$). In~\cite{Grimmer:2014kx}, the SF to CB phase transition at half-filling happens around $V/t \sim 4.6$, while with density-induced hopping, this transition happens around $\gamma\sim 0.053$ (equivalent to $V/t\sim 3.27$). We were not able to resolve the nature of the CB-SF phase transition at $n=0.5$. We did not detect a supersolid phase neither found evidence of a first-order phase transition (we changed the interaction strength $\gamma$  in increments of $ 0.5 \% $).
For $\gamma > 0.053$, upon doping with particles or holes from half-filling, we enter the CBSS phase. Here, diagonal long-range order and off-diagonal long-range order coexist as shown from a finite superfluid density $\rho_s$ and a finite structure factor $S(\pi, \pi)$. For large enough doping, on both particle and hole sides, the supersolid disappears in favor of a SF phase via a second-order phase transition. 

In Fig.~\ref{FIG2}(b) we show the phase diagram at tilt angle $\theta=11.25^{\circ}$. The qualitative shape of the phase diagram is the same as in Fig.~\ref{FIG2}(a) but with a slightly more extended SF region. Here too, the density-induced hopping parameters are  negative ($T_x\sim-0.15$ and $T_y\sim-0.17$). At this angle, the repulsive interaction along the $x$ direction has decreased, while the repulsive interaction along the $y$ direction does not change significantly. This leads to a slightly larger superfluid region compared to the $\theta=0^{\circ}$ phase diagram. We investigated the SF-CB transition at filling factor $n=0.5$ and found hysteresis curves as a function of the interaction strength $\gamma$ for the superfluid density $\rho_s$ and structure factor $S(\pi, \pi)$. These hysteresis curves signal a  first-order phase transition.

There exists a  qualitative change in the phase diagrams in going from  $\theta=11.25^{\circ}$ to $\theta=45^{\circ}$. This change happens at $\theta\sim30^{\circ}$, where the solid phase and the supersolid phase change from the CB pattern to SS pattern. From Ref~\cite{Zhang:ws}, we know that there exists an emulsion phase at $\theta \sim30^{\circ}$ which is challenging to resolve numerically.  Phase diagrams $\theta\lesssim30^{\circ}$ have a similar shape to the one at  $\theta=11.25^{\circ}$, while phase diagrams for $30^{\circ}\lesssim\theta\lesssim58^{\circ}$ have a similar shape to the one at $\theta=45^{\circ}$. For larger tilt angles, phase diagrams are similar to the one for $\theta=67.5^{\circ}$.

\begin{figure}[h]
\centering
\includegraphics[trim=1.2cm 1.2cm 1.2cm 1.2cm, clip=true, width=0.48\textwidth]{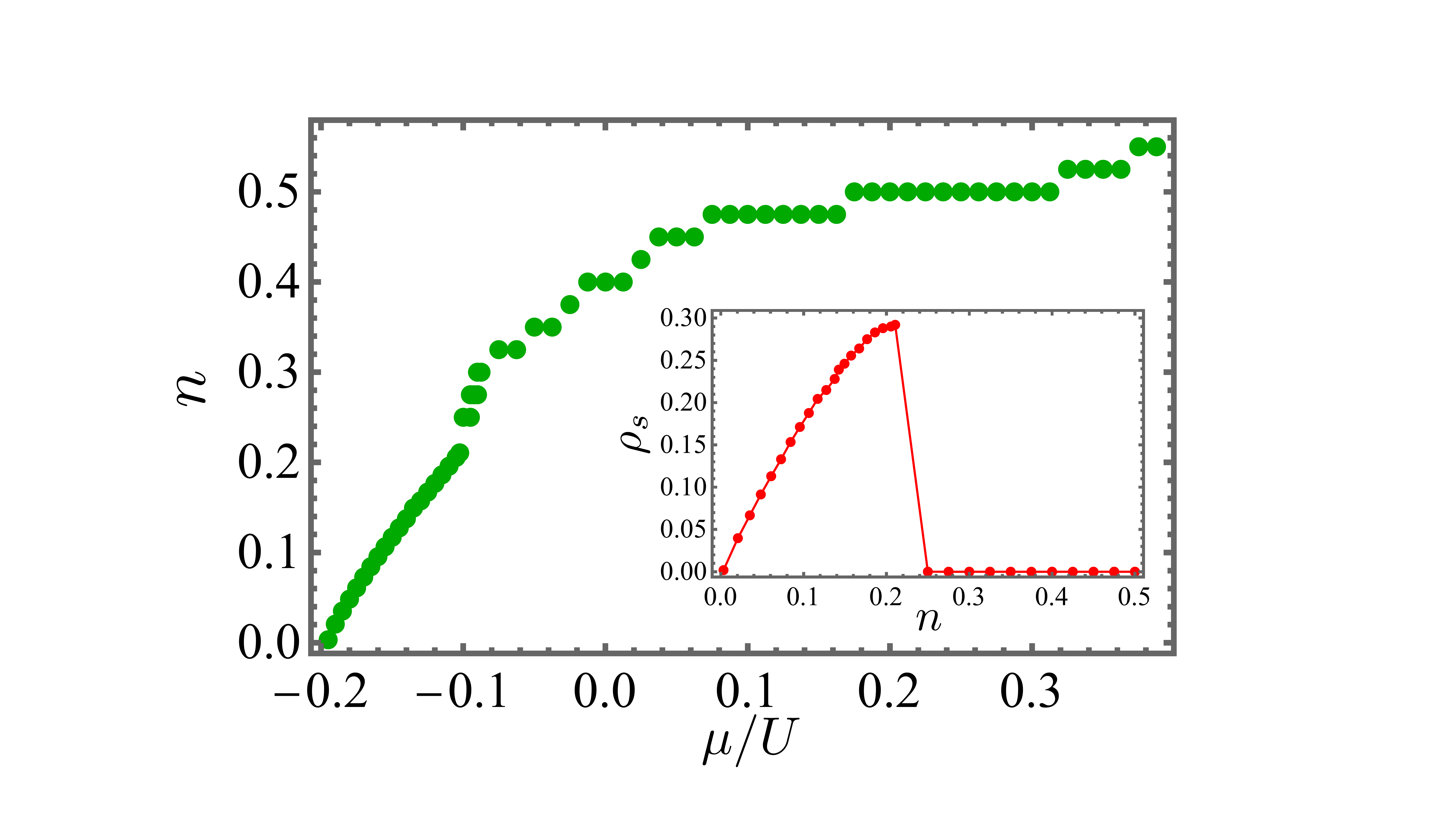}
\caption{ Plots correspond to parameters $U/t=20$, $\theta=45^{\circ}$, $\gamma=0.081$, and $L=40$. Main plot : filling factor $n$ as a function of $\mu/U$. Inset: superfluid density $\rho_s$ as a function of filling factor $n$. 
When not visible error bars are within symbol size.}
\label{FIG3}
\end{figure}

Fig.~\ref{FIG2}(c) shows the ground state phase diagram at tilt angle $\theta=45^{\circ}$. At this angle, the density-induced hopping along the $y$-direction $T_y$ is negative while $T_x$ is positive with $0.135<|T_y|/t<0.291$ and $0.149<T_x/t<0.210$ for the range of $\gamma$ considered. The dipolar interaction along the $x$ axis is attractive stabilizing a stripe solid phase at filling factor $n=0.5$ and $\gamma\gtrsim0.0564$. The density induced hopping has little effect on the onset of the SS. To check this, we ran simulations with no density induced hopping and found the onset of the SS at $\gamma \sim0.059$.
We have studied the transition from SF to SS at half filling and observed that a supersolid intervenes in between within a narrow range, $0.0534 < \gamma < 0.0564$. For $0.0564<\gamma<0.068$, a SSS phase also appears upon doping the stripe solid with particles or holes. For low enough doping, the solid order of the SSS is the same as for the SS. For larger doping, instead, we see that stripes are not uniformly spaced (we will discuss this below in more details for the incompressile phase). Interestingly, as $\gamma$ further increases, the SSS phase disappears in favor of a succession of incompressible ground states stabilized at rational filling factors (this succession will become dense in the thermodynamic limit), similar to the classical devil's staircase~\cite{Hubbard:1978im,Fisher:1980be,Bak:1982ev}.  This is seen in Fig.~\ref{FIG3} where filling factor $n$ is plotted as a function of chemical potential $\mu/U$ at $L=40$, $\theta=45^{\circ}$, $U/t=20$, and $\gamma\sim0.081$. When $\mu/U>-0.10$, one can observe several plateaus at different rational filling factors. 
These plateaus correspond to incompressible ground states. For all filling factors considered, we have observed a stripe phase similar to the one at n=0.5, the only difference being that the spacing between stripes changes with filling factor and can be irregular to accommodate a certain filling. The inset shows the superfluid density $\rho_s$ as a function of filling factor $n$. At lower filling factor, the SF density is finite but goes to zero abruptly at $n\sim 0.25$. This abrupt change in $\rho_s$ indicates that the SF phase disappears in favor of the incompressible phase (IP) through a first-order phase transition as confirmed from hysteretic behavior in the $n$ vs. $\mu/U$ curve (not shown here).  We mark the first-order phase transition with the blue solid region which corresponds to density range for which one would observe phase coexistence. Finally,  we notice that we did not find any staggered SF, expected for larger $|T_y|$ and/or larger filling factor~\cite{Kraus:2020es}.


\begin{figure}[h]
\centering
\includegraphics[trim=6cm 0cm 12cm 0cm, clip=true, width=0.65\textwidth]{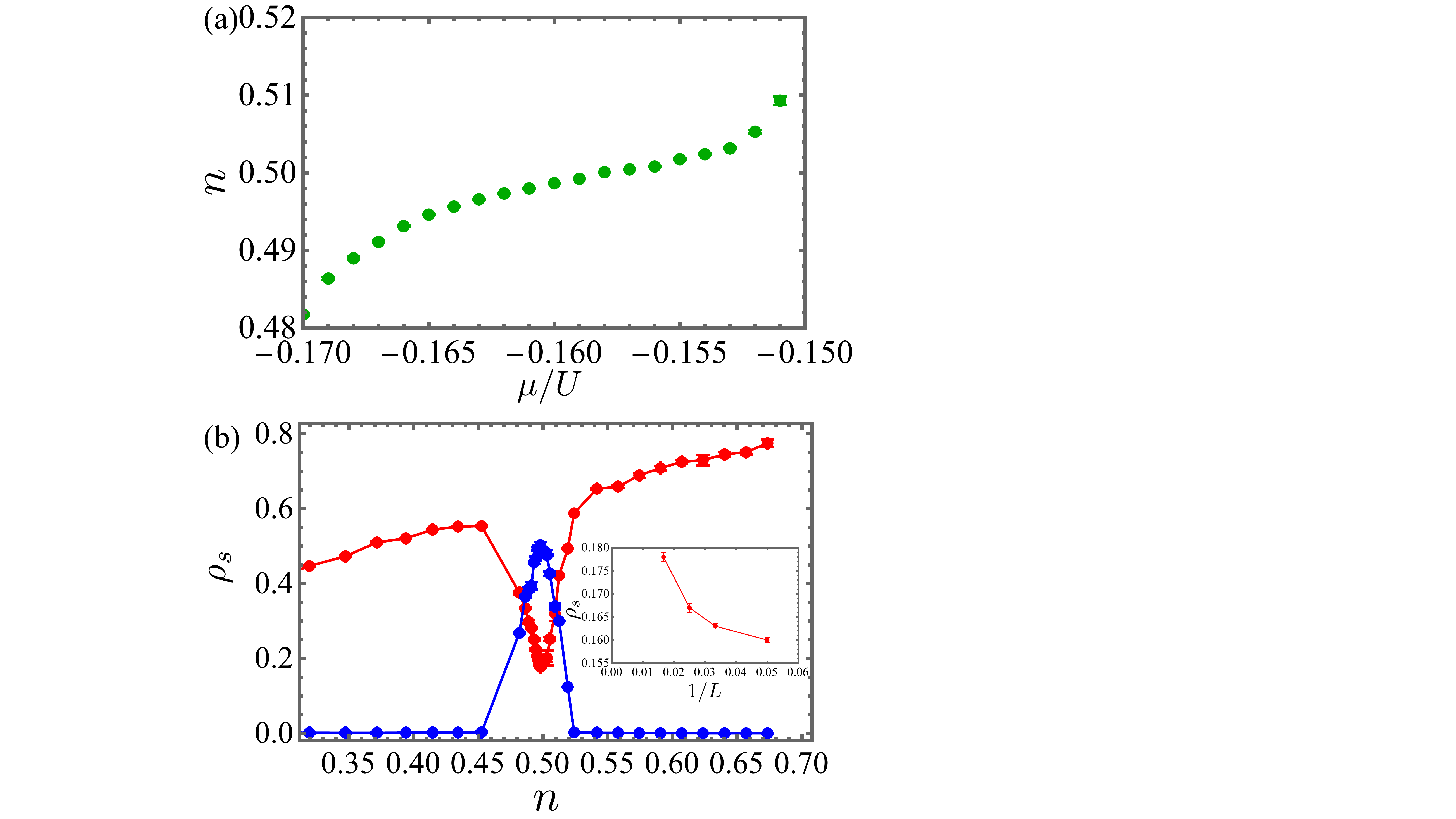}
\caption{
Main plots correspond to parameters $U/t=20$, $\theta=67.5^{\circ}$, $\gamma=0.0361$, and $L=60$. (a) filling factor $n$ as a function of $\mu/U$. (b) superfluid density $\rho_s$ and structure factor $S(0,\pi)$ as a function of filling factor $n$. Inset shows the superfluid density $\rho_s$ as a function of $1/L$  at filling factor $n=0.5$. When not visible error bars are within symbol size.}
\label{FIG4}
\end{figure}
The shape of the ground state phase diagram changes when $\theta \gtrsim 58^{\circ}$. 
Fig~\ref{FIG2}(d) shows the phase diagram at tilt angle $\theta=67.5^{\circ}$. At this angle, the density-induced hopping along the $y$-direction $T_y$ is negative while $T_x$ is positive with $0.113<|T_y|/t<0.127$ and $0.204<T_x/t<0.215$ for the range of $\gamma$ considered. The phase diagram features a SS realized at half filling and for $\gamma\gtrsim 0.0363$. Upon doping the SS with particles, we find a SSS for all values of $\gamma$ considered. On the hole side, the situation is more complex. A SSS is only stabilized at lower $\gamma$ values, while at larger $\gamma$ values we find an incompressible phase.  Interestingly, we find that incompressible ground states at lower filling factor can be stabilized for values of $\gamma$ smaller than what required to stabilize the stripe solid at half filling. This incompressible phase is realized at fractional filling factors, in analogy with what discussed at  $\theta=45^{\circ}$. For lower values of dipolar interaction $\gamma$, as the density is increased, the incompressible phase disappears in favor of a SF via a first-order phase transition (shaded green region) as confirmed by hysteretic behavior in the $n$ vs. $\mu$ curve (not shown here). Upon further increasing the density, one enters the SSS phase which, interestingly, survives also at $n=0.5$ within the range $0.0358\lesssim\gamma\lesssim0.0363$. Evidence of the supersolid at half filling is shown in Fig.~\ref{FIG4}. In Fig.~\ref{FIG4}(a) we plot the filling factor $n$ as a function of chemical potential $\mu/U$ at $L=60$, $\theta=67.5^{\circ}$, $U/t=20$, and $\gamma=0.0361$. Notice that there does not exist a plateau at $n=0.5$. This excludes the existence of an incompressible phase at half filling for this value of $\gamma$.
In Fig.~\ref{FIG4}(b), we plot superfluid density $\rho_s$ and structure factor $S(0, \pi)$ as a function of filling factor $n$. One can clearly see the coexistence of superfluidity and solid order at $n=0.5$. Here, a finite $\rho_s$ persists (and increases) as the system size is increased, as shown in the inset of Fig.~\ref{FIG4}(b) where we plot the superfluid density as a function of $1/L$ at $n=0.5$. For $\gamma \gtrsim 0.0363$, superfluid order disappears and a SS is stabilized at $n=0.5$.
We notice that, in the parameter region considered, we have not found any evidence of staggered SF.

\section{Experimental Realization}
\label{sec:sec5}
Various ultracold bosonic systems with different particle species are capable to explore the phase diagrams proposed above. These systems include atoms with magnetic dipole moments such as Cr
\cite{Griesmaier:2005fd,Naylor:2015bs}, Er \cite{Aikawa:2012ic,Aikawa:2014if,Baier:2016ga}, and Dy \cite{Lu:2012bd,Lu:2011hl}, polar bimolecules such as Er$_{2}$ \cite{Frisch:2015gm}, KRb \cite{Yan:2013fna}, NaK \cite{Seesselberg:2018ff}, and atoms in Rydberg states \cite{Booth:2015fp, Schauss:2015ch}. The two-dimensional system can be realized by loading BEC ensembles into optical lattices formed by overlapping three perpendicularly crossed laser beams with their retro-reflected beams. While the trap depths along two dimensions are equal, the trap depth along the third dimension should be much larger to keep the lattice system two-dimensional or quasi-two-dimensional. The orientation of the electric or magnetic dipole moments can be adjusted freely using external electric or magnetic fields and the value of $\gamma$ depends on lattice constant, external fields, gas species, and which states they are in. The filling factor $n$ can be tuned by changing trap depth and onsite interactions through Feshbach field. The Feshbach resonance also makes the $a_{s}/a$ ratios proposed in this work all accessible. When the lattice constant equals 532 nm, for Cr which has a $\gamma$ of $\sim0.06$, all phases can be stabilized with appropriate choice of filling factor and tilt angle. For Er, Dy which have larger  dipole moments (Er: $\gamma\sim0.27$, Dy: $\gamma\sim0.53$), different quantum phases can be realized under different filling conditions at any tilt angle. If the lattice constant changes from 532 nm to 266 nm, the $\gamma$ values above are changed by a factor of 2. When using ultracold polar molecules, even larger dipole moments, correspondingly $\gamma$, can be obtained. For example, ultracold polar molecule Er$_{2}$ can give a $\gamma$ as high as $\sim6.20$ hence the entire phase diagram can be explored.

Various detection methods are capable to detect the phases arising under different conditions. The above discussed systems with different tilt angles are in the SF phase when $\gamma$ is small. The SF phase can be detected by observing interference patterns in the time-of-flight imaging of the ultracold quantum gases released from traps \cite{Greiner:2002es, RevModPhys.80.885}. Several other quantum phases emerge after increasing $\gamma$. These phases pose challenges to time-of-flight detection. However, ultracold quantum gases in these phases have modulated density distributions in lattices, for example, SS and SSS have periodical density modulations in optical lattices while CB has particles distributed with a checkboard pattern in lattices. These patterns can be directly observed using state of the art quantum gas microscopes with single-site-resolved imaging capacity~\cite{sherson2010, simon2011,yang2021siteresolved}. The IP phase which has a modulated density distribution can also be feasibly observed using quantum gas microscopes.

\section{Conclusion}
\label{sec:sec6}
We have studied the ground states of soft-core dipolar bosons with density-induced hopping as described by the extended Bose-Hubbard model on a square lattice. Dipoles are tilted in the $xz$ plane. The parameters entering the effective model are calculated starting from the parameters that can be tuned experimentally, e.g. scattering length and dipole moment which both contribute to the onsite interaction, long-range interaction, and strength of density-induced hopping. We have found the ground state phase diagrams of this system at tilt angles $\theta=0^{\circ}$, $11.25^{\circ}$, $45^{\circ}$, and $67.5^{\circ}$. We have observed that, as the dipolar interaction increases, the superfluid phase at half filling factor is destroyed in favor of either a checkerboard or stripe solid phase for tilt angle $\theta \lesssim 30^{\circ}$ or $\theta \gtrsim 30^{\circ}$ respectively. At tilt angles $\theta \gtrsim 58^{\circ}$, we have found that, as the dipolar interaction strength increases, solid phases first appear at filling factor lower than $0.5$. For $\theta=45^{\circ}$ and $67.5^{\circ}$, we have observed the presence of a supersolid phase intervenes between the superfluid and stripe solid phase at half filling. All the phases discussed here can be realized experimentally with magnetic atoms or polar molecules.

{\textit{Acknowledgements}}  
We would like to thank Giovanna Morigi, Jacub Zakrzewski, Rebecca Kraus, and Peter Schauss for fruitful discussions.
The computing for this project was performed at the OU Supercomputing Center for Education $\&$ Research (OSCER) at the University of Oklahoma (OU) and the cluster at Clark University.

\appendix 
\section{Calculation of Hamiltonian Parameters} \label{apdx:calparams}
\begin{figure*}
  \centering
    \includegraphics[width=0.98\textwidth]{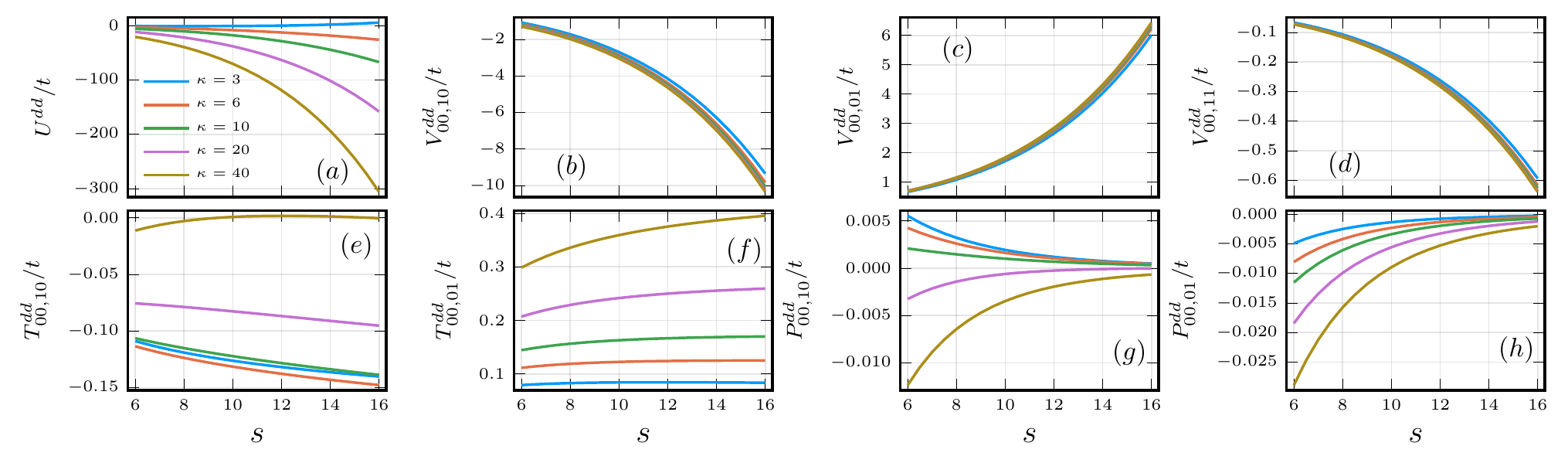}
    \caption{\label{fig:bandparamsuvtp}(Color online) Dipolar contribution to Hamiltonian parameters as a function of lattice depth $s$ for $\kappa = 3, 6, 10, 20, 40$ at tilt angle $\theta = 67.5^{\circ}$ and $\gamma = 1/\pi^3$. We plot (a) onsite interaction $U^{dd}/t$; (b) nearest-neighbor interaction in $\hat{\mathbf{x}}$ direction $V^{dd}_{\mathbf{i}, \mathbf{i}+\hat{\mathbf{x}}}/t$; (c) nearest-neighbor interaction in $\hat{\mathbf{y}}$ direction $V^{dd}_{\mathbf{i}, \mathbf{i}+\hat{\mathbf{y}}}/t$; (d) next-nearest-neighbor interaction in $\hat{\mathbf{x}}+\hat{\mathbf{y}}$ direction $V^{dd}_{\mathbf{i}, \mathbf{i}+\hat{\mathbf{x}}+\hat{\mathbf{y}}}/t$; (e) density-induced tunneling in $\hat{\mathbf{x}}$ direction $T^{dd}_{\mathbf{i}, \mathbf{i}+\hat{\mathbf{x}}}/t$; (f) density-induced tunneling in $\hat{\mathbf{y}}$ direction $T^{dd}_{\mathbf{i}, \mathbf{i}+\hat{\mathbf{y}}}/t$; (g) pair tunneling in $\hat{\mathbf{x}}$ direction $P^{dd}_{\mathbf{i}, \mathbf{i}+\hat{\mathbf{x}}}/t$; (h) pair tunneling in $\hat{\mathbf{y}}$ direction $P^{dd}_{\mathbf{i}, \mathbf{i}+\hat{\mathbf{y}}}/t$. }
  \end{figure*} 

In a separable lattice potential, the Wannier function can be written as $W(x,y,z) = w(x)w(y)w(z)$. $w(x)$ is the one dimensional Wannier function in a $1$D optical lattice. In the lattice coordinates, $\mathbf{r} \rightarrow \mathbf{r} / a$, $w(z) = (\pi \kappa)^{1/4} exp(-\pi^2 \kappa z^2 / 2)$ is the ground state wavefunction of the harmonic trap in $z$ direction. The contribution to the parameters of the Bose-Hubbard model can be calculated separately for contact interaction and dipole-dipole interaction. Labeling the sites in the square lattice as $\mathbf{i} = (i_x, i_y)$, the general interaction comes from integrals of four Wannier functions at sites $\mathbf{i}, \mathbf{j}, \mathbf{k}, \mathbf{l}$. In units of the recoil energy, the contact interaction gives
\begin{eqnarray}
\label{eq:contactintegral}
\nonumber U^c_{\mathbf{i}, \mathbf{j}, \mathbf{k}, \mathbf{l}} &=& \frac{8a_s}{\pi a} \int d\mathbf{r} W^*_{\mathbf{i}}(\mathbf{r})W^*_{\mathbf{j}}(\mathbf{r})W_{\mathbf{k}}(\mathbf{r})W_{\mathbf{l}}(\mathbf{r}) \\ 
&=& \frac{8a_s}{\pi a} \sqrt{\frac{\kappa \pi}{2}} \int dx dy \mathcal{W}^*_{\mathbf{i}}\mathcal{W}^*_{\mathbf{j}}\mathcal{W}_{\mathbf{k}}\mathcal{W}_{\mathbf{l}},
\end{eqnarray}
where $\mathcal{W}_{\mathbf{i}} = w_{i_x}(x)w_{i_y}(y)$ is the two dimensional Wannier function. Then the contribution to the Hamiltonian parameters from contact interaction are $U^c = U^c_{\mathbf{i i i i}}, V^c_{i j} = U^c_{\mathbf{i j i j}} + U^c_{\mathbf{i j j i}} = 2U^c_{\mathbf{i j i j}}, T^c_{\mathbf{i j}} = -(U^c_{\mathbf{i i i j}} + U^c_{\mathbf{i i j i}}) / 2 = - U^c_{\mathbf{i i i j}}, P^c_{\mathbf{i j}} = U^c_{\mathbf{i i j j}}$. Typical values of this contribution are $U^c / t = 30.5, V^c / t = 0.006, T^c / t = 0.104, P^c / t = 0.003$ for $a_s / a = 0.014$, corresponding to $a_s = 100 a_0$ for $\ce{^{87}Rb}$ at lattice spacing $a = 377$ nm~\cite{Luhmann:2012kq, Best:2009gr}.\\

The contributions from dipole-dipole interactions can be calculated by Fourier transform
\begin{eqnarray}
\label{eq:ddintegral}
\nonumber D_{\mathbf{i}, \mathbf{j}, \mathbf{k}, \mathbf{l}} &=& \int d\mathbf{r} d\mathbf{r}^\prime W^*_{\mathbf{i}}(\mathbf{r})W^*_{\mathbf{j}}(\mathbf{r^\prime})V_d(\mathbf{r}^\prime-\mathbf{r})W_{\mathbf{k}}(\mathbf{r^\prime})W_{\mathbf{l}}(\mathbf{r}) \\ 
&=& \frac{1}{(2\pi)^3} \int d\mathbf{k} \tilde{W}_{\mathbf{il}}(-\mathbf{k}) \tilde{V}_d(\mathbf{k}) \tilde{W}_{\mathbf{jk}}(\mathbf{k}),
\end{eqnarray}
where the Fourier transform of the product of two Wannier functions with the same coordinate $\mathbf{r}$ is
\begin{eqnarray}
\nonumber \tilde{W}_{\mathbf{il}}(\mathbf{k}) &=& \int d \mathbf{r} W^*_{\mathbf{i}}(\mathbf{r}) W_{\mathbf{l}}(\mathbf{r}) e^{-i \mathbf{k} \cdot \mathbf{r}} \\ 
\nonumber &=& e^{-\frac{k_z^2}{4\pi^2\kappa}} \int dx dy \mathcal{W}^*_{\mathbf{i}}(x,y) \mathcal{W}_{\mathbf{l}}(x,y) e^{-i(k_x x + k_y y)} \\
\end{eqnarray}
The Fourier transform of dipolar interaction reads
\begin{eqnarray}
\tilde{V}_d(\mathbf{k}) = 4\pi \gamma \left(cos^2(\beta) - \frac{1}{3}\right),
\end{eqnarray}
where $\beta$ is the angle between $\mathbf{k}$ and the dipole moment. We can further integrate out $k_z$,
\begin{eqnarray}
\nonumber D_{\mathbf{i}, \mathbf{j}, \mathbf{k}, \mathbf{l}} &=& \frac{1}{(2\pi)^2}\int d \mathbf{k}_{xy} \tilde{\mathcal{W}}_{\mathbf{il}}(-\mathbf{k}_{xy}) \tilde{\mathcal{V}}(\mathbf{k}_{xy}) \tilde{\mathcal{W}}_{\mathbf{jk}}(\mathbf{k}_{xy}) \\
\end{eqnarray}
and the effective two dimensional interaction is
\begin{eqnarray}
\nonumber && \tilde{\mathcal{V}}(\mathbf{k}_{xy}) = 2\pi \gamma \Bigg\{ \sqrt{2\pi \kappa}\left( n_z^2 - \frac{1}{3} \right) + \bigg[ \frac{\left( n_x k_x + n_y k_y\right)^2}{\sqrt{k_x^2+k_y^2}} \\ && - n_z^2 \sqrt{k_x^2+k_y^2}\bigg] e^{\frac{k_x^2+k_y^2}{2\pi^2\kappa}} Erfc\left[\sqrt{\frac{k_x^2+k_y^2}{2\pi^2\kappa}}\right] \Bigg\}
\end{eqnarray}
where $\mathbf{n} = (n_x, n_y, n_z)$ is the unit vector along the dipole moments, and $Erfc(x)$ is the complementary error function. The dipolar part of Hamiltonian parameters are $U^{dd} = D_{\mathbf{i i i i}}, V^{dd}_{\mathbf{i j}} = D_{\mathbf{i j i j}} + D_{\mathbf{i j j i}}, T^{dd}_{\mathbf{i j}} = D_{\mathbf{i i i j}}, P^{dd}_{\mathbf{i j}} = D_{\mathbf{i i j j}}$. 

Some results for tilt angle $\theta = 67.5^{\circ}$ and $\gamma = 1/\pi^3$ are depicted in Fig.~\ref{fig:bandparamsuvtp}, where we show the dipolar contribution to Hamiltonian parameters as a function of lattice depth $s$ for $\kappa = 3, 6, 10, 20, 40$. As the lattice depth $s$ is increased, the dipolar contribution to onsite interaction increases from negative to positive for small $\kappa = 3$. For larger $\kappa$ instead, it becomes more negative by increasing either $s$ or $\kappa$. These behaviors are opposite to those at small tilt angles. The nearest-neighbor interaction in $\hat{\mathbf{x}}$ direction $V^{dd}_{\mathbf{i}, \mathbf{i}+\hat{\mathbf{x}}}$ and the next-nearest-neighbor interaction in $\hat{\mathbf{x}}+\hat{\mathbf{y}}$ direction $V^{dd}_{\mathbf{i}, \mathbf{i}+\hat{\mathbf{x}}+\hat{\mathbf{y}}}$ are negative as expected, and become more negative by increasing either $s$ or $\kappa$, while the nearest-neighbor interaction in $\hat{\mathbf{y}}$ direction $V^{dd}_{\mathbf{i}, \mathbf{i}+\hat{\mathbf{y}}}$ behaves just the opposite. The dependence of off-site interactions on $\kappa$ becomes negligible for large enough $\kappa$ because we are approaching a perfect 2D lattice. The density-induced tunneling in $\hat{\mathbf{x}}$ direction is negative and becomes more negative by increasing $s$ for $\kappa \le 20$. If we increase $\kappa$ with fixed $s$, it becomes more negative first but  then increases and goes towards positive values. 
In the $\hat{\mathbf{y}}$ direction, it becomes more positive by either increasing $s$ or increasing $\kappa$. The dependence on $s$ is very small for small $\kappa$. We notice that, at small angles, the density-induced tunneling goes from positive to negative as we increase $\kappa$. The pair tunneling is very small compared to other parameters, so we neglect it in the Hamiltonian.


\bibliography{induce}

\end{document}